\documentstyle[11pt]{article}
\def\half{\mbox{$\frac{1}{2}$}}     

\begin{document}
\special{vc:  ssvcid simplgr.tex 1.3 Fri 95Aug18  17:32 \quad
              ssvcid TeX'd \today \quad
              vcidtag
}


%
\title{Precis of General Relativity%
\footnote{Remarks presented at the Army Research Office workshop on
the Global Positioning System, 2--3 August 1995, Chapel Hill NC}
\footnote{UMCP-PP-96-16; gr-qc/9508043}
}
\author{   \sc
            Charles W. Misner\\
           \em
            Department of Physics, University of Maryland
           \\ \em
            College Park MD 20742-4111 USA\\
           \rm
         e-mail: \tt misner@umail.umd.edu
        }
\date{3 August 1995}
\maketitle

\begin{abstract}
    Omitting the motivations and historical connections, and also
the detailed calculations, I state succinctly the principles that
determine the relativistic idealization of a GPS system.  These
determine the results that Ashby presents in his tutorial.
\end{abstract}

\section*{}
    A method for making sure that the relativity effects are
specified correctly (according to Einstein's General Relativity) can
be described rather briefly.  It agrees with Ashby's approach but
omits all discussion of how, historically or logically, this
viewpoint was developed.  It also omits all the detailed
calculations.  It is merely a statement of principles.

    One first banishes the idea of an ``observer''. This idea aided
Einstein in building special relativity but it is confusing and
ambiguous in general relativity.  Instead one divides the theoretical
landscape into two categories.  One category is the
mathematical/conceptual model of whatever is happening that merits
our attention.  The other category is measuring instruments and the
data tables they provide.

    For GPS the measuring instruments can be taken to be either ideal
SI atomic clocks in trajectories determined by known forces, or else
electromagnetic signals describing the state of the clock that
radiates the signal.  Each clock maintains its own proper time (but
may convert this via software into other information when it
transmits).  We simplify to assume it transmits its own proper time
without random or systematic errors, so that its increments $d\tau_T$
are simple physical data.  Any other clock receiving these signals
can record data tables showing the increments $d\tau_R$ in the SI
proper time of the clock at the receiver corresponding to differences
$d\tau_T$ in the proper times encoded in the signal it receives from
some other identified transmitting clock.  Once conventional zeros of
time are identified for each clock, each transmitting and receiving
pair produces a data table $\tau_R(\tau_T)$.  These segments of data
are to be reproduced by computations from the conceptual model, with
any residuals understood on the basis of expected sources of noise
and unmodelled phenomena.

    A user ``fix'' or relativistic ``event'' is the simultaneous
reception of signals from four GPS satellites, or its equivalent from
short extrapolations from nearly simultaneous signals. This user may
not have a reliable clock but should be able to determine the time
and position of the event from knowledge of the proper times encoded
in the received signals, the identities of the transmitters, and the
mathematical/conceptual model that defines the meaning of time and
position for this purpose.  System software aims to make the user
calculations standard and practical, with many of the computational
results encoded in the transmitted signals.

    What is the conceptual model?  It is built from Einstein's
General Relativity which asserts that spacetime is curved.  This
means that there is no precise intuitive significance for time and
position.  [Think of a Caesarian general hoping to locate an outpost.
Would he understand that 600 miles North of Rome and 600 miles West
could be a different spot depending on whether one measured North
before West or visa versa?]  But one can draw a spacetime map and
give unambiguous interpretations. [On a Mercator projection of the
Earth, one minute of latitude is one nautical mile everywhere, but
the distance between minute tics varies over the map and must be
taken into account when reading off both NS and EW distances.] There
is no single best way to draw the spacetime map, but unambiguous
choices can be made and communicated, as with the Mercator choice for
describing the Earth.

    The conceptual model for a relativistic system is a spacetime map
or diagram plus some rules for its interpretation.  For GPS the
attached Figure is a simplified version of the map.  The real
spacetime map is a computer program that assigns map locations $xyzt$
to a variety of events.  In the Figure the $t$ time axis is vertical,
and two of the three space $xyz$ axes are suggested horizontally. The
wide center swath is the Earth which occupies the same location,
centered on the central axis of the map, at all times.  Marked on the
surface of the Earth is a long spiral representing, e.g., a clock at
USNO. The position of this clock as the Earth rotates is described by
the coordinates of this curve on the (corresponding conceptual four
dimensional) map, $x(t), y(t), z(t)$ where $xyzt$ are distances
measured by a Euclidean ruler on the (conceptual four dimensional)
graph paper parallel to its axes.  The scale factors needed for
interpreting this spacetime map are provided by the metric.  In the
map projection (coordinate system) from which the GPS model starts
(an Earth Centered Inertial coordinate system, ECI) the metric is
    \begin{equation}\label{e-ECI}
     {d\tau}^2 = [1 + 2(V - \Phi_0)/c^2]dt^2
                    - [1 - 2V/c^2]
                      (dx^2 + dy^2 + dz^2)/c^2 \quad .
    \end{equation}
    Here $V$ is the Newtonian gravitational potential of the
Earth, approximately
    \begin{equation}\label{e-phi}
     V = - (GM/r) [1 - \half J_2 (R/r)^2 (3 \cos^2 \theta -1)] \quad .
    \end{equation}
    The constant $\Phi_0$ is chosen so that a standard SI clock ``on
the geoid'' (e.g., USNO were it at sea level) would give, inserting
its world line $x(t), y(t), z(t)$ into equation~(\ref{e-ECI}), just
$d\tau = dt$ where $d\tau$ is the physical proper time reading of the
clock.  It is a theorem that if this choice is made for one clock on
the geoid it applies to all.

    Equation~(\ref{e-ECI}) defines not only the gravitational field
that is assumed, but also the coordinate system in which it is
presented.  There is no other source of information about the
coordinates apart from the expression for the metric.  It is also not
possible to define the coordinate system unambiguously in any way that
does not require a unique expression for the metric.  In most cases
where the coordinates are chosen for computational convenience, the
expression for the metric is the most efficient way to communicate
clearly the choice of coordinates that is being made.  Mere words
such as ``Earth Centered Inertial coordinates'' are ambiguous unless
by convention they are understood to designate a particular
expression for the metric, such as equation~(\ref{e-ECI}).

    Using equation~(\ref{e-ECI}) one can place tic marks along the
world line of any clock to show changes in its proper time (which are
to be physical changes directly displayed and transmitted by the
clock). The computation is just to insert the clock trajectory $x(t),
y(t), z(t)$ to find $d\tau$ from equation~(\ref{e-ECI}) as a thus
specified multiple of $dt$.  This applies both to Earth fixed clocks,
to satellite clocks, and to clocks with any other motion $x(t), y(t),
z(t)$ that has been incorporated in the map.  The ``map'' here means
a computer program that is designed to produce the trajectories
$x(t), y(t), z(t)$ of each modelled object.

    The rules for drawing clock world lines or trajectories on the
spacetime map (in the computer program) are simplest for dragfree
satellites and for electromagnetic signals in vacuum.  In these cases
the world line must be a (timelike, resp.\ lightlike) geodesic of the
metric~(\ref{e-ECI}), i.e., a solution of an ordinary differential
equation constructed using the coefficients (scale factors) in
equation~(\ref{e-ECI}).  The electromagnetic signals have the special
property that their trajectories also satisfy $d\tau^2 = 0$ in
equation~(\ref{e-ECI}).  By finding a lightlike geodesic that connects
one tic mark $\tau_T$ on one clock world line to another mark
$\tau_R$ on another clock, the map shows how one entry in the physical
data table $\tau_R(\tau_T)$ is computed in the mathematical model.
Once the observed data tables are being reproduced adequately in the
mathematical model, its assignments of $xyzt$ coordinates to events
identify the time and position of those events.

    In sum, the $txyz$ time and position values provided by GPS are
not simple physical times and positions.  Physical times and
positions exist but, due to spacetime curvature, cannot be naturally
associated with quadruples of numbers.  Physical times and positions
are identified on a spacetime map by their $xyzt$ map coordinates
which depend on the ``projection'' (coordinate system) chosen in
designing that particular map.  The ECI map defined by
equation~(\ref{e-ECI}) is the simplest to describe.  More practical
maps have been defined in which the space coordinates of geodetic
benchmarks on Earth are nearly constant and change only due to
tectonic and volcanic activity.  To identify such an Earth fixed
coordinate system one gives these coordinates as specified functions
of those used in the ECI metric.  This results in a metric
expression different from equation~(\ref{e-ECI}) and allows results
computed in the ECI coordinate system to be reported in the second
coordinate system.

\section*{Figure}
This spacetime diagram shows the Earth, a fixed location (USNO) on
the rotating Earth, a satellite orbiting the Earth, and an
electromagnetic (EM) signal propagating from an event T on the
satellite's world line to an event R on the USNO world line.  Two of
the three $xyz$ space axes are indicated.  The $t$ time axis is at
the center of the Earth.  Any point on this diagram or map can be
located by its $xyzt$ coordinates which are measured along the
coordinate axes as conventional Cartesian coordinates for points
(events) on this map.  To deduce physical separations between
(nearby) points on the map one must use equation~(\ref{e-ECI}) to
convert the separations $dx\,dy\,dz\,dt$ read from the map into
a physically measurable proper time interval $d\tau$.

\section*{References}
Neil Ashby, ``A tutorial on Relativistic Effects in the Global
Positioning System'', NIST Contract No.40RANB9B8112, February 1990.

Neil  Ashby, ``Relativistic Effects in the Global Positioning
System'', NIST Contract No.40RANB9B8112, August 1995.

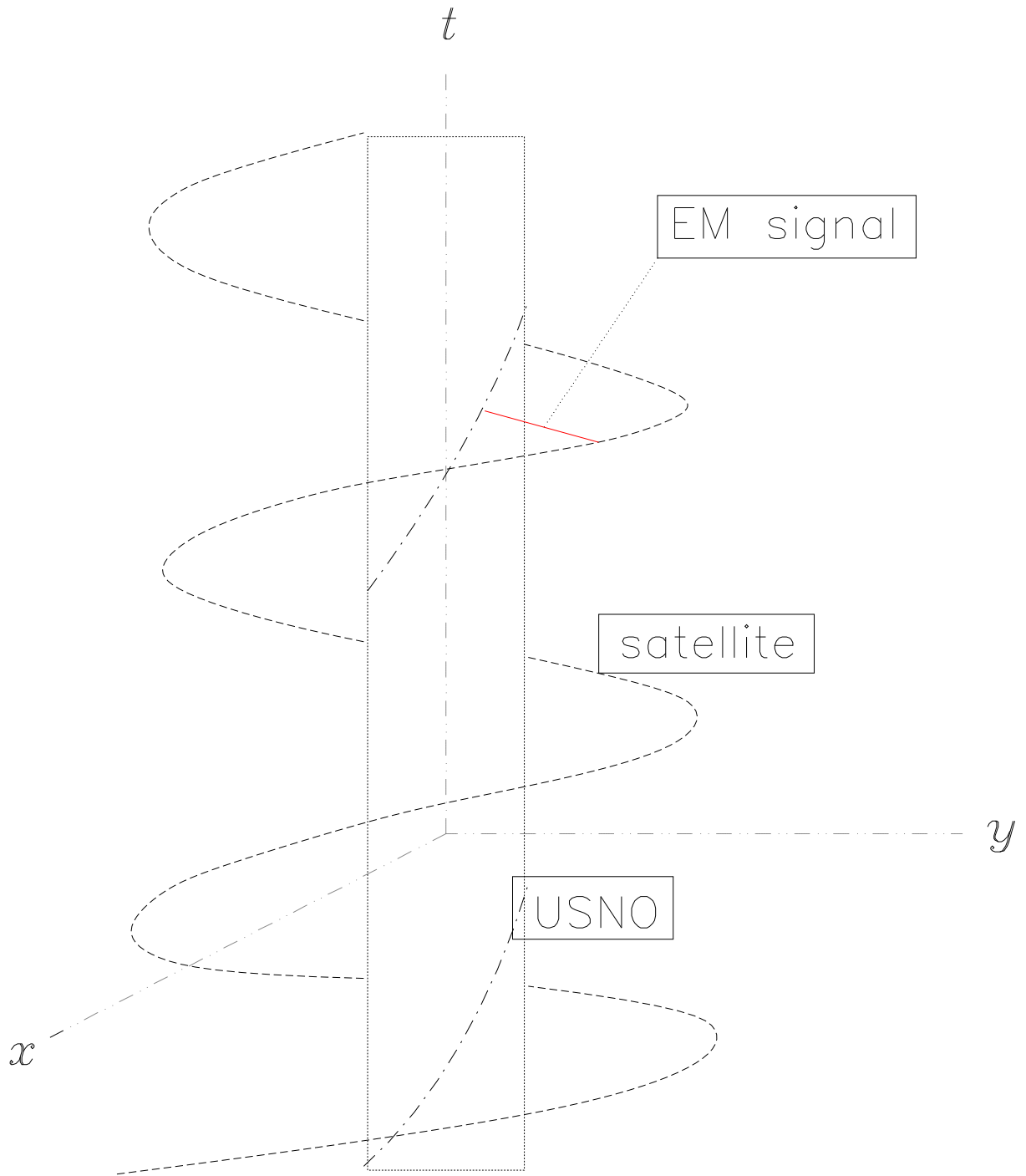
\end{document}